\documentclass[twocolumn,prb,superscriptaddress,showpacs,citeautoscript,floatfix]{revtex4}

\usepackage{graphicx}
\usepackage{dcolumn}   
\usepackage{bm}        
\usepackage{flafter}
\usepackage{color}
\usepackage{hyperref}

\graphicspath{{eps+pdf/}}

\begin{document}
\title{Structural and optical investigations of the iron-chalcogenide superconductor Fe$_{1.03}$Se$_{0.5}$Te$_{0.5}$ under high pressure}

\author{Pallavi S. Malavi}
\affiliation{High Pressure and Synchrotron Radiation Physics Division,
Bhabha Atomic Research Centre, Trombay, Mumbai 400085, India}

\author{S. Karmakar}
\email[E-mail:~]{sdak@barc.gov.in}
\affiliation{High Pressure and Synchrotron Radiation Physics Division,
Bhabha Atomic Research Centre, Trombay, Mumbai 400085, India}

\author{N. N. Patel}
\affiliation{High Pressure and Synchrotron Radiation Physics Division,
Bhabha Atomic Research Centre, Trombay, Mumbai 400085, India}

\author{Surinder M. Sharma}
\affiliation{High Pressure and Synchrotron Radiation Physics Division,
Bhabha Atomic Research Centre, Trombay, Mumbai 400085, India}

\date{\today}

\begin{abstract}

Iron-chalcogenide superconductor Fe$_{1.03}$Se$_{0.5}$Te$_{0.5}$ has been investigated under high pressure using synchrotron based x-ray diffraction and mid-infrared reflectance measurements at room temperature. Pressure dependence of the superconducting transition temperature (T$_c$) of the same sample has been determined by temperature-dependent resistance measurements up to 10 GPa. Although the high pressure orthorhombic phase ($\textit{Pbnm}$) starts emerging at 4 GPa, structural transition becomes clearly observable above 10 GPa. A strong correlation is observed between the Fe(Se,Te)$_{4}$ tetrahedral deformation in the tetragonal phase ($\textit{P4/nmm}$) and the sharp rise of T$_c$ up to $\sim$4 GPa, above which T$_c$ is found to be almost pressure independent at least up to 10 GPa. A subtle structural modification of the tetragonal phase is noticed above 10 GPa, suggesting a structural transition with possible Fe$^{2+}$ spin-state transition. The evolution with pressure of the optical conductivity shows that the Drude term increases systematically with pressure up to 5.4 GPa, indicating the evolution of the tetragonal phase towards a conventional metallic state. At further higher pressures the Drude term reduces drastically implying a poor metallic character of the high pressure orthorhombic phase. Our results suggest that occurrence of large onset T$_c$ above 4 GPa is due to a systematic increase of high pressure orthorhombic phase fraction.

\end{abstract}



%

\maketitle

\section{Introduction}

Discovery of superconductivity in iron-chalcogenides, with the layered anti-PbO tetragonal structure at room temperature, has raised considerable interest in studying these compounds in more detail because of their simplest crystal structure among various iron-based superconductors~\cite{Hsu2008}. Interestingly, although $\alpha$-FeSe exhibits spin-fluctuation driven superconducting ground state (with T$_c$ $\sim$9K), unlike other iron-based superconductors (FeSCs), at ambient pressure no long range magnetic order is established at low temperature~\cite{Imai2009}. However moderate amount of external pressure drives the system into a state where static magnetic order is realized~\cite{Bendele2010}. Structurally, due to the absence of alkali ions or charge-reservoir layers between the Fe$_{2}$Se$_{2}$ superconducting layers, the system is found to be very much compressible and a strong correlation is noticed between the structural parameters and superconducting transition temperature.  Numerous high pressure investigations have been carried out on undoped $\alpha$-FeSe to elucidate the large pressure dependence of its superconducting transition temperature~\cite{Mizuguchi2008,Miyoshi2009,Huang2009,Okabe2010,Uhoya2012,Kumar2010,Garbarino2009,Braithwaite2009,Margadonna2009,Medvedev2009}. Several structural studies indicate that both the room temperature tetragonal phase (S.G. $\textit{P4/nmm}$, z=2) and the low temperature (superconducting) orthorhombic phase (S.G. $\textit{Cmma}$, z=4) undergo structural transition to an orthorhombic (S.G. $\textit{Pbnm}$, z=4) phase~\cite{Kumar2010,Garbarino2009,Braithwaite2009,Uhoya2012}. There are also reports of pressure-induced structural transition to $\beta$-FeSe phase (S.G. $\textit{P6$_{3}$/mmc}$)~\cite{Margadonna2009,Medvedev2009} and amorphization~\cite{Stemshorn2010}.

Recent observation of the structural distortion upon Te-doping in $\alpha$-FeSe compounds and the correlated change in the superconducting transition temperature T$_c$~\cite{Yeh2008,Horigane2009} has stimulated interest in the detailed investigations of the structural and electronic properties of this compound under high pressure. Several high pressure investigations have been performed on optimally Te-doped FeSe and the superconducting T$_c$ is found to rise sharply with pressure up to $\sim$3 GPa, reaching T$_{cmax}$ $\sim$23K~\cite{Gresty2009,Huang2009,Gooch2012,Tsoi2009,Karmakar2013}. Two different observations are reported on low temperature structural stability of Te-doped FeSe compounds under high pressure, viz., transition to a distorted high pressure structure (monoclinic symmetry) above 3 GPa and amorphization above 11.5 GPa~\cite{Gresty2009,Stemshorn2009}, lacking a complete structural evolution results. Various theoretical investigations have predicted the chalcogen height dependent change in the magnetic ordering and the electronic properties in Te-doped FeSe compounds~\cite{Moon2010,Kumar2012,Ciechan2012}. It is therefore imperative to perform detailed high pressure experimental investigations on the structural and optical properties to understand the effect of pressure on the electronic structural changes of the pressurized compound. This will, in effect, help understand the behavior of T$_c$ as a function of pressure and also to make a comparison this with the results of undoped FeSe.

In this work we examine the detailed high pressure structural evolution at room temperature of the superconducting compound Fe$_{1.03}$Se$_{0.5}$Te$_{0.5}$ and corresponding changes in its frequency dependent optical conductivity. Synchrotron based x-ray diffraction and mid-infrared (mid-IR) reflectance measurements have been performed under high pressures up to 19 GPa and 15 GPa respectively. Tetragonal $\textit{P4/nmm}$ structure starts transforming to $\textit{Pbnm}$ orthorhombic phase at a very low pressure (4 GPa), but the transition becomes prominent only above $\sim$10 GPa. Dramatic changes in the low frequency (Drude) spectral weight of the optical conductivity are noticed above this pressure. The results have been discussed with the observed pressure dependence of superconducting T$_c$, obtained from temperature dependent (2-200K) resistance measurements up to 10 GPa.


\section{Experimental Details}

Polycrystalline Fe$_{1.03}$Se$_{0.5}$Te$_{0.5}$ has been prepared by the solid state reaction method. Stoichiometric amounts of Fe (99.9$\%$), Se (99.5$\%$) and Te (99.5$\%$) are mixed, and the pressed powder of this mixture are sealed in an evacuated quartz tube and heated at 700$^\circ$C for 24 hr. Structural characterization by x-ray diffraction measurements on as-grown sample shows a mixture of tetragonal $P4/nmm$ (60$\%$) and hexagonal $P6_{3}/mmc$ phase (40$\%$). No trace of excess Fe was detected. Magnetic susceptibility measurement was performed using a SQUID magnetometer (Quantum Design) under an external magnetic field of 10 Oe and the onset superconducting T$_c$ was found to be $\sim$13.5 K.

Pressure-induced structural evolution has been studied using a diamond anvil cell (DAC) by monochromatic ($\lambda$= 0.68881 ${\AA}$) x-ray powder diffraction measurements at the XRD1 beamline of Elettra Synchrotron facility. Methanol-Ethanol-Water (MEW 16:3:1) mixture is used as pressure transmitting medium (PTM) to provide nearly hydrostatic conditions and gold is used as pressure marker in x-ray diffraction measurements~\cite{Dewaele2004}. Two dimensional diffraction patterns were recorded on a Dectris Pilatus2M detector and these were converted to one dimensional diffraction profiles using FIT2D software~\cite{Hammersley1998}. The structural parameters are determined by means of GSAS Rietveld refinement analysis program~\cite{Larson1986}.

Mid-IR reflectance spectra under high pressures have been collected at the SISSI beamline of Elettra synchrotron facility, with the help of an IR microscope (Hyperion) equipped with a liquid nitrogen cooled HgCdTe (MCT) detector and coupled to a Bruker IFS66v interferometer. DAC was mounted with type IIa synthetic diamond anvils for this measurement. A thin pallet of finely ground sample was placed in the gasket hole, which was then filled with KCl as pressure medium. Reflectance at the sample-diamond interface were measured in the wide frequency range (700 - 12000 cm$^{-1}$). At each pressure, the intensity of light ($I_{s}(\omega)$) reflected from the sample surface (within DAC) is measured and normalized with the light intensity ($I_{0}(\omega)$) reflected from the external face of diamond window, to obtain a quantity $r(\omega)=\frac{I_s(\omega)}{I_0(\omega)}$. At the end of the pressure run, another measurement is performed on the open diamond surface ($I_{d}(\omega)$,$I_{0}(\omega)$) by removing the sample. The ratio $R(\omega)=\frac{I_d(\omega)}{I_0(\omega)}$ is assumed to be pressure independent. The reflectance $R_{s-d}(\omega)$ at the sample-diamond interface is thus obtained from the equation $R_{s-d}(\omega)=\frac{r(\omega)}{R(\omega)}R_{dia}$, where R$_{dia}$=[$\frac{n_{dia}-1}{n_{dia}+1}$]$^{2}$ and $n_{dia}$ being the real refractive index of diamond in air. For this experiment, pressure inside the DAC was monitored with the $\textit{in-situ}$ ruby luminescence method~\cite{Mao1986}.

The pressure evolution of the superconducting T$_c$ has been measured on the same batch sample up to $\sim$10 GPa using a miniature DAC (MPI-Stuttgart version) and an optical cryostat (temperature range 2 - 300K). A quasi-four probe resistance measurement technique with $\textit{in-situ}$ ruby pressure measurement was employed. The details of the DAC preparation, resistance measurement and sample pressure determination technique have been described elsewhere~\cite{Karmakar2013}. A great care was taken in order to obtain the sample pressure near T$_c$ and resistance with high accuracy. Resistance is measured on a tiny sample (a pressed pellet of Fe$_{1.03}$Se$_{0.5}$Te$_{0.5}$, diameter $\sim$50 $\mu$m) using 10$\mu$m thin gold electrical leads, that remains in good contact with the sample by the applied pressure through the solid PTM. Finely powdered NaCl (soft solid, remains insulating through out this temperature and pressure range) has been used as the PTM.

\section{Results and Discussions}

\subsection{X-ray Diffraction}

\begin{figure}[tb]
\centerline{\includegraphics[width=75mm,clip]{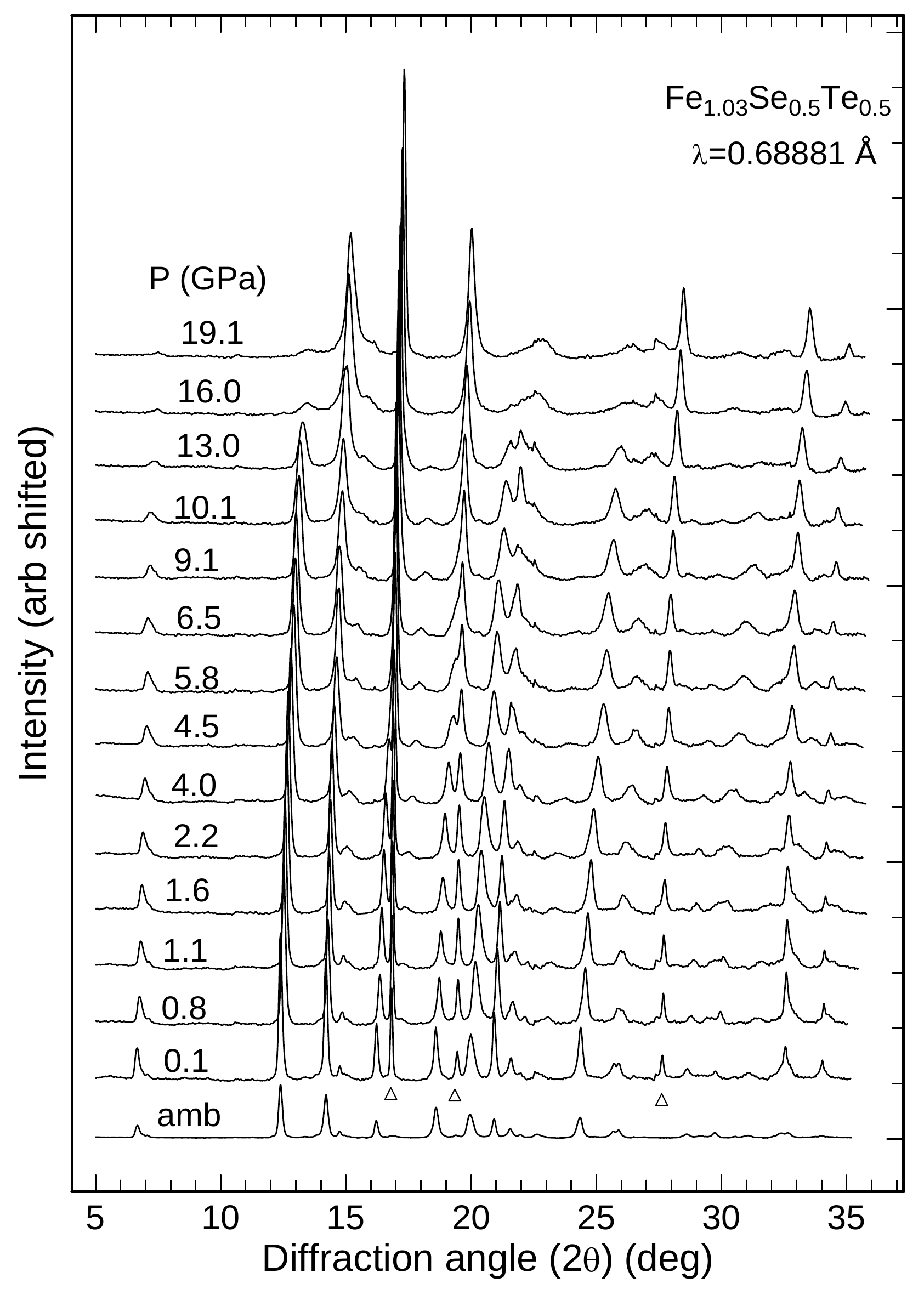}}
\caption{\label{fig1} X-ray diffraction patterns at ambient and various high pressures at room temperature.  Diffraction peaks marked with up triangles are the Au (pressure marker) Bragg peaks.}
\end{figure}

\begin{figure}[tb]
\centerline{\includegraphics[width=85mm,clip]{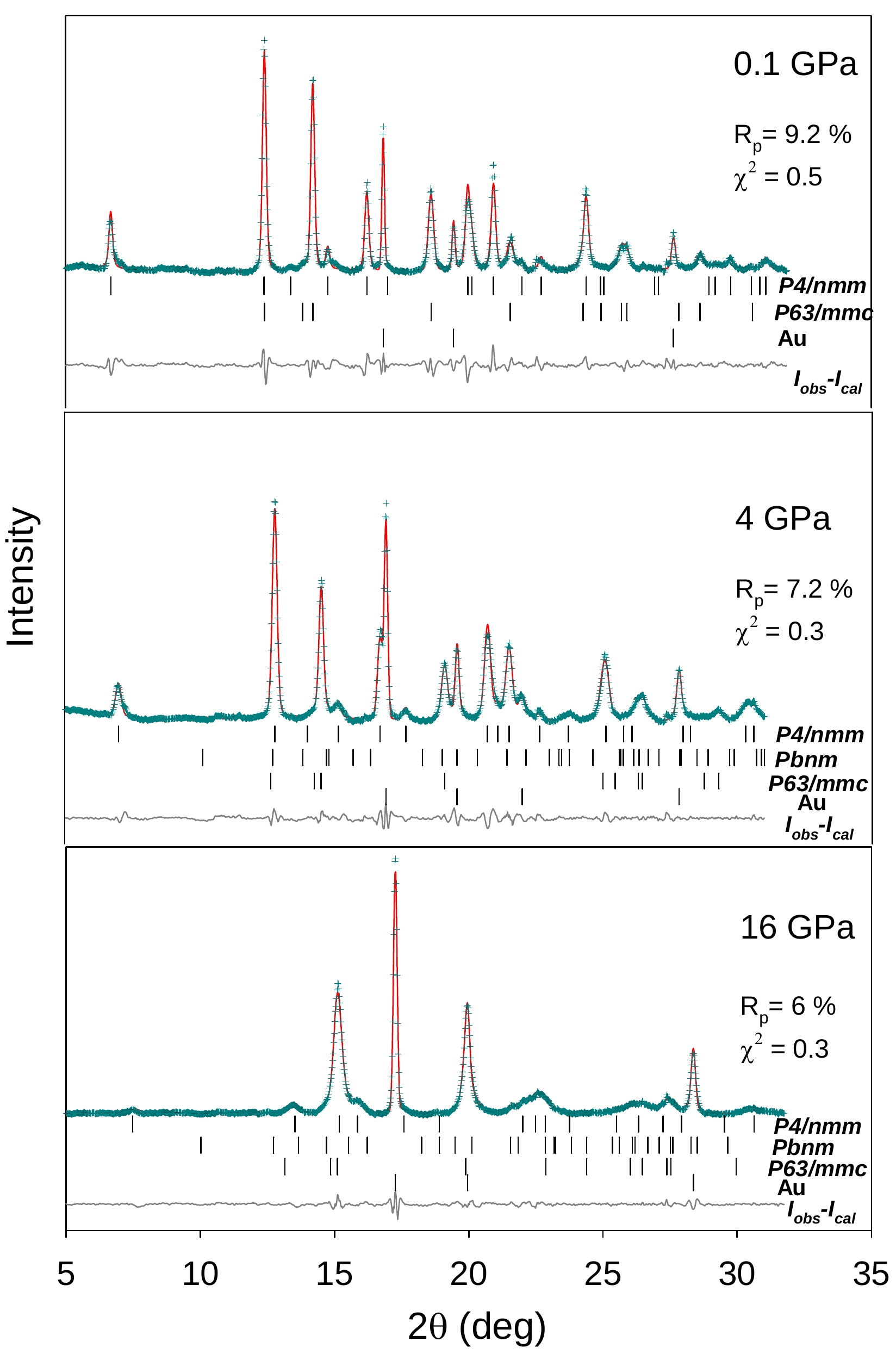}}
\caption{\label{fig1} (Color online) Rietveld refinement profiles at three different pressures (0.1, 4.0 and 16 GPa). Cross hair symbols represent the experimental data, and solid lines are calculated diffraction pattern. The phase markers and difference plots are shown below.}
\end{figure}


\begin{figure}[tb]
\centerline{\includegraphics[width=90mm,clip]{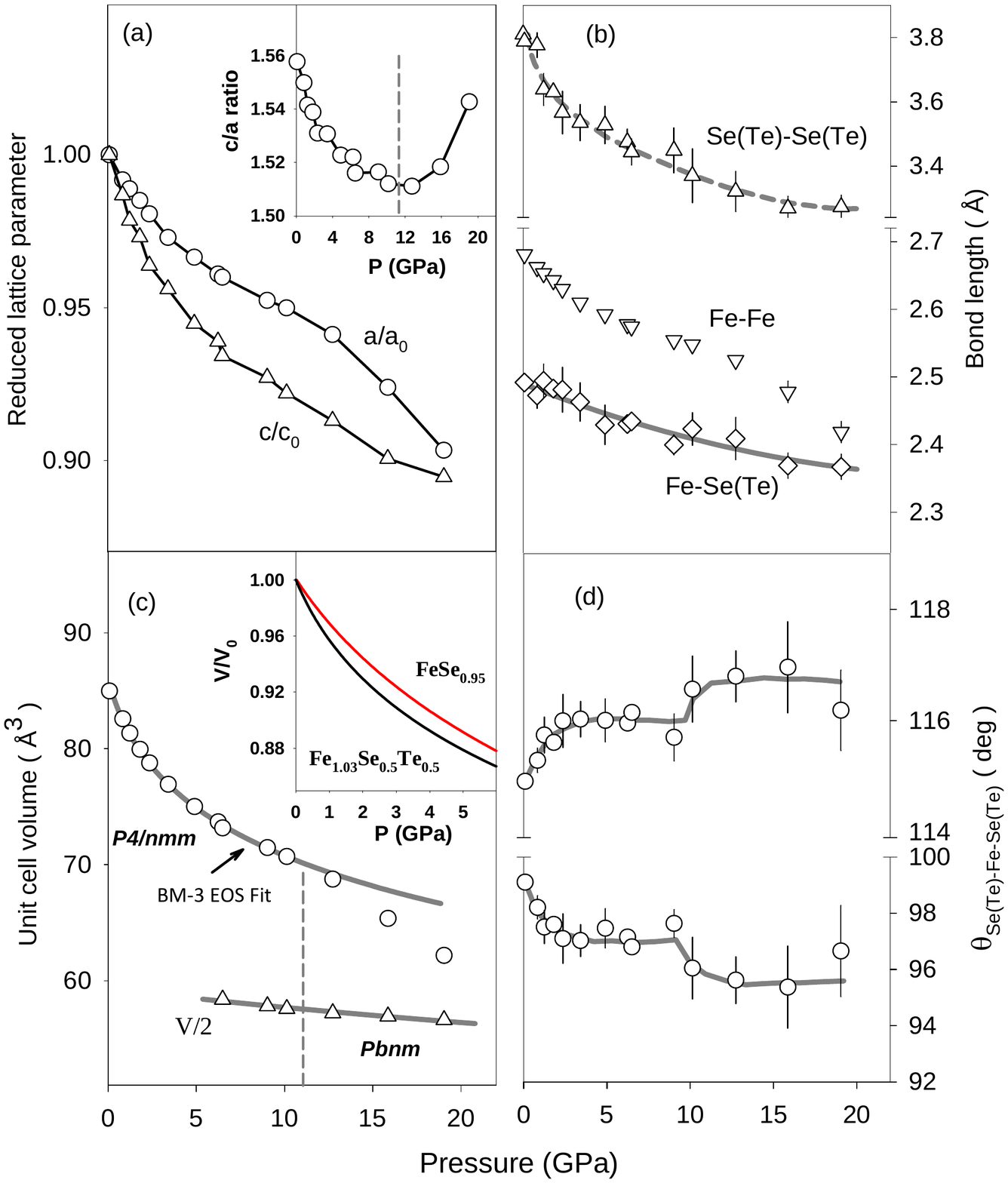}}
\caption{\label{fig2} Pressure-induced structural changes in room temperature tetragonal phase (S.G. $P4/nmm$) (a)Reduced lattice parameters (a/a$_{0}$, c/c$_{0}$) as a function of pressure; inset, pressure variation of $c/a$ ratio (b) Changes of interatomic distances  [Fe-Fe and Fe-Se(Te) bonds and non-bonded Se(Te)-Se(Te)], (c) unit cell volume as a function of pressure, 3rd order Birch-Murnaghan equation of state, up triangles represent the half of the cell volume of the $Pbnm$ phase for comparison with tetragonal phase. Inset, higher compressibility of Fe$_{1.03}$Se$_{0.5}$Te$_{0.5}$ compared to FeSe (from ref. 10) is shown. (d) Tetrahedral angles Se(Te)-Fe-Se(Te) as a function of pressure.}
\end{figure}

Diffraction patterns of Fe$_{1.03}$Se$_{0.5}$Te$_{0.5}$ at various high pressures are shown in Figure 1. Low pressure diffraction profiles are analyzed considering a mixture of tetragonal ($\textit{P4/nmm}$) and hexagonal ($\textit{P6$_{3}$/mmc}$) phases. The tetragonal phase being responsible for the low temperature superconductivity, we have followed its structural evolution under pressure. At ambient pressure tetragonal structure lattice parameters are a=3.7923(3)${\AA}$ and c=5.9083(5)${\AA}$. The diffraction patterns remain almost unchanged up to 10 GPa, above which the diffraction peaks corresponding to the tetragonal phase reduce in intensity and some additional peaks (for example, at $2\theta$= 16.1$^\circ$ and 22$^\circ$) appear at the shoulder of the diffraction peaks of the hexagonal phase, indicating a structural transition to a phase other than the existing hexagonal phase. The additional peaks very well agree with the orthorhombic (S.G. $\textit{Pbnm}$, z=4) phase, already reported as the high pressure phase of undoped-FeSe~\cite{Kumar2010,Garbarino2009,Braithwaite2009,Uhoya2012}. However, unlike undoped FeSe, structural transition of tetragonal Fe$_{1.03}$Se$_{0.5}$Te$_{0.5}$ is rather sluggish. Although new diffraction peaks are clearly observable at 10 GPa and above, Rietveld structural analysis indicate that the orthorhombic phase starts emerging even at 4 GPa. Figure 2 shows the Rietveld refinement profiles at three different pressures (0.1, 4 and 16 GPa). The phase fractions of the three phases ($P4/nmm, Pbnm, P6_{3}/mmc$) are 56$\%$, 4$\%$ and 40$\%$ respectively at 4 GPa and 18$\%$, 40$\%$ and 42$\%$ respectively at 16 GPa. This indicates that the molar fraction of the hexagonal phase remains unchanged (within the accuracy limit of GSAS analysis) at high pressure. The lattice parameters of the high pressure orthorhombic $\textit{Pbnm}$ phase at 16 GPa are a=5.7932(8)${\AA}$, b=5.3848(12)${\AA}$ and c=3.6819(5)${\AA}$.

In Figure 3, we have shown in detail the pressure-induced evolution of structural parameters of the tetragonal phase.  Large anisotropic compressibility (more compressible along c-axis) is noticed due to the presence of a van der Waal interlayer gap between the Fe$_{2}$(Se,Te)$_{2}$ layers. The $c/a$ ratio decreases systematically up to $\sim$10 GPa above which this ratio starts increasing. The concomitant increase in compressibility at this pressure is reflected in the plot of the pressure variation of the unit cell volume. So, the bulk compressibility of the remaining tetragonal phase changes from its initial value at $\sim$10 GPa. The pressure versus unit cell volume data up to $\sim$10 GPa, when fitted to 3rd order Birch-Murnaghan equation of state (EOS)~\cite{Birch1978}, gives zero pressure bulk modulus as B=18 GPa and its pressure derivative B'=13. Tetragonal phase of Fe$_{1.03}$Se$_{0.5}$Te$_{0.5}$ is therefore found to have the smallest bulk modulus in any FeSCs (twice as compressible (soft) as the undoped $\alpha$-FeSe, as can be seen from our comparison made in inset in Fig 3c ~\cite{Garbarino2009,Margadonna2009}). A somewhat higher value of bulk modulus reported by Stemshorn et al.~\cite{Stemshorn2009}, is probably due to an improper fitting of EOS in the low pressure range. Pressure-induced changes of the interatomic distances are plotted in Fig 3b. The interlayer non-bonded Se(Te)-Se(Te) distance and the Fe-Se(Te) bond distance decrease monotonically up to the highest pressure of this investigation. A rapid change in the Fe-Fe bond distance (same as lattice parameter $a$) is noticed above 10 GPa. The Fe(Se,Te)$_{4}$ tetrahedra is further elongated under pressure and the pressure variation of the Se(Te)-Fe-Se(Te) angles are plotted in figure 3d.

Structural refinement shows that the average anion height ($z_{Se,Te}$) of the $P4/nmm$ phase rapidly increases with pressure up to $\sim$3 GPa and remains pressure independent at higher pressures till $\sim$10 GPa, where the anion height rises abruptly (as seen in Figure 5a). A similar feature has earlier been reported for undoped $\alpha$-FeSe~\cite{Kumar2010,Okabe2010}. A monotonic increase of anion heights in the tetragonal FeSe$_{0.5}$Te$_{0.5}$ has recently been predicted by Ciechan et.al.~\cite{Ciechan2012}. The observed changes of anion heights are also reflected in the pressure-induced FeSe(Te)$_{4}$ tetrahedral deformation (2zc/(a/$\sqrt{2}$)) (Figure 4b). The observed sudden enhancement of the tetrahedral elongation and the anomalous changes of $c/a$ ratio and the axial compressibility indicates a structural transition to a new tetragonal phase above 10 GPa. A larger bulk compressibility of this phase may be attributed to the possible pressure-induced electron delocalization or the spin crossover, as observed in updoped-FeSe \cite{Kumar2011}. Tetrahedral elongation changes the electronic structure of the system, specially the Fe$^{2+}$ orbital character \cite{Kumar2012,Tamai2010}. In the low pressure region, with increasing pressure the electronic structural changes drive the system into more metallic character (enhanced Drude spectral weight, as discussed below), enhancing effective carrier concentration, favoring a higher T$_c$~\cite{Schilling2006}. In figure 4c is plotted the pressure dependence of the observed superconducting T$_c$ (as discussed in next section) and a strong structural correlation with the superconductivity is noticed.

\subsection{Resistance measurements}

\begin{figure}[tb]
\centerline{\includegraphics[width=75mm,clip]{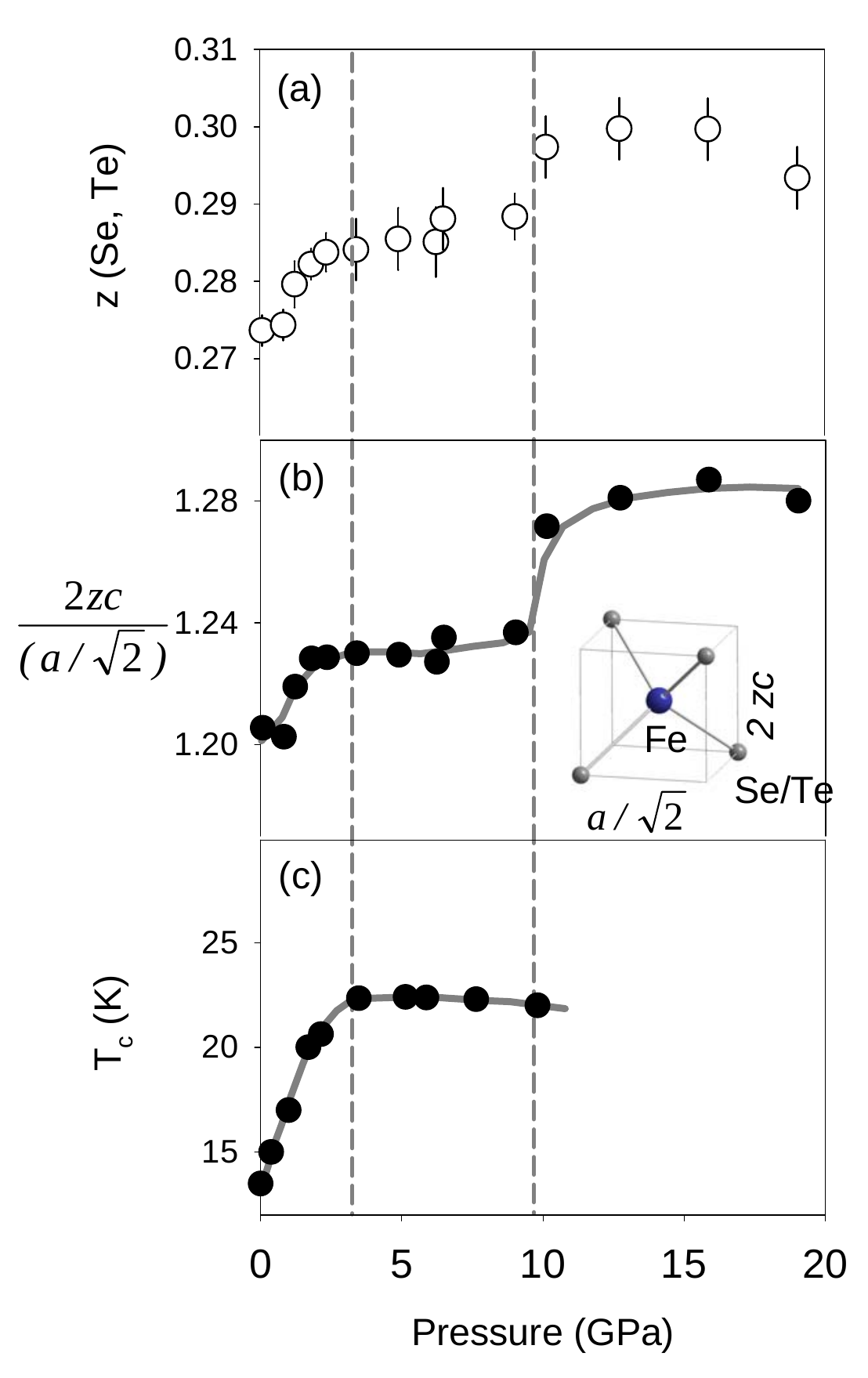}}
\caption{\label{fig2} (a) pressure variation of the average anion (Se/Te) height $z$, (b) pressure induced tetrahedral deformation, denoted by $2zc/(a/\sqrt{2})$. Inset, Fe(Se,Te)$_{4}$ tetrahedral dimension (c) The superconducting transition temperature, T$_c$, as a function of pressure. Solid lines are guide to eye.}
\end{figure}

\begin{figure}[tb]
\centerline{\includegraphics[width=75mm,clip]{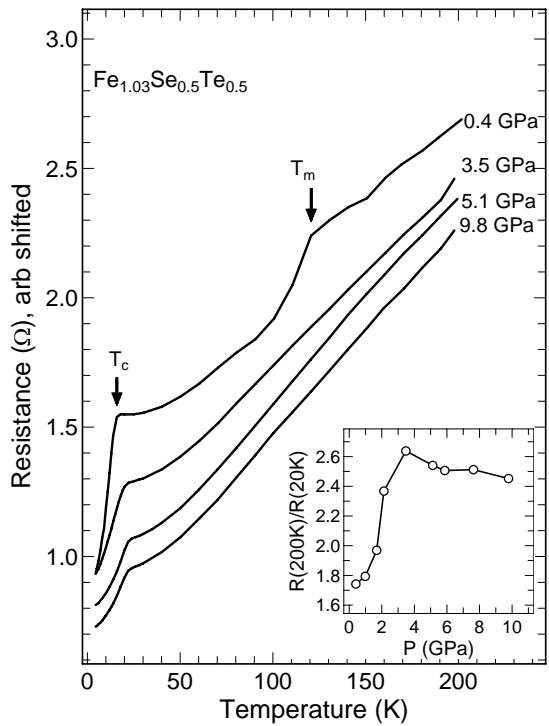}}
\caption{\label{fig2} Temperature-dependent resistance of Fe$_{1.03}$Se$_{0.5}$Te$_{0.5}$ at a few representative pressures. T$_c$ indicates the superconducting transition temperature. At low pressure, resistance anomaly due to SDW-like magnetic transition is indicated by T$_m$. Inset, residual resistance, $R(200K)/R(20K)$, as a function of pressure.}
\end{figure}

Figure 5 shows the temperature dependent resistance at a few representative pressures. Offsets are given to the $R(T)$ data at various pressures for the sake of clarity. At low pressures, two characteristic features are observed. A resistance anomaly due to spin-density wave (SDW)-type transition is observed at $\sim$120K and upon increasing pressure this transition temperature ($T_m$) decreases and the feature disappears above 1 GPa. It should be noted that we observe, for the first time, a clear picture of the SDW-like resistance anomaly in Te-doped FeSe compound. Te-doping increases the average anion height and therefore affects the band structure and Fermi surface significantly, establishing the static magnetic order even at ambient pressure~\cite{Kumar2012}. However, like many other FeSCs, application of pressure enhances the spin fluctuation causing reduction of SDW feature upon increasing pressure~\cite{Imai2009}.

At further lower temperature, superconducting transition is observed by a sudden drop of the resistance (a non-zero resistance in the superconducting state is due to the lead/contact probe resistance in the quasi-four probe measurements). The onset T$_c$ at each pressure is determined by the intersection of the two quadratic fits to the data above and below the point of inflection. Our results show that T$_c$ increases with pressure with a large initial slope and reaches its maximum (T$_{cmax}$ $\sim$22.4K) near 4 GPa, indicating a typical unconventional nature (spin fluctuation driven) of superconductivity, but unlike other reports ~\cite{Tsoi2009,Gresty2009}, T$_c$ reduces only marginally beyond this pressure. The pressure dependence of T$_c$ at zero pressure is found to be 4.74 K/GPa, agreeing very well with a previous study~\cite{Yeh2008} and on strained FeSe$_{0.5}$Te$_{0.5}$ thin film~\cite{Gooch2012}. Although Fe$_{1.03}$Se$_{0.5}$Te$_{0.5}$ is found to have slightly higher initial pressure dependence of T$_c$ than undoped FeSe ($\sim$4.3K/GPa,~\cite{Braithwaite2009}), the relative volume derivative $dlnT_c/dlnV = -B(dlnT_c/dP)$ (a parameter of T$_c$ dependence on structural modification) is found to be only half of that for undoped FeSe. The residual resistance $R(T = 200K)/R(T = 20K)$ is found to increase systematically with pressure up to 4 GPa (inset of Figure 5), suggesting an enhanced metallic behavior in normal state at higher pressures.

\subsection{Infrared Reflectance}

\begin{figure}[tb]
\centerline{\includegraphics[width=75mm,clip]{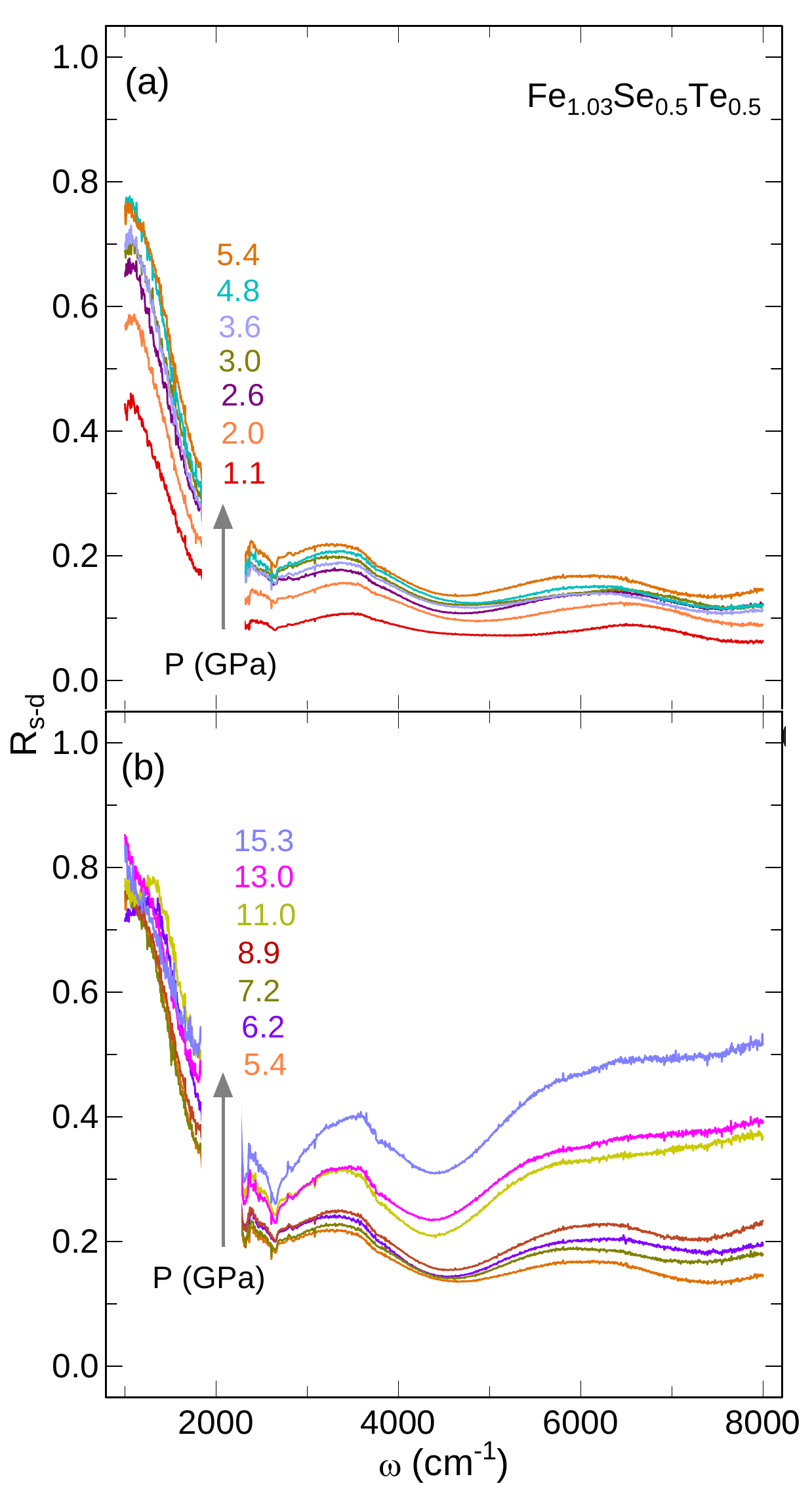}}
\caption{\label{fig5} (Color online) Reflectivity spectra of the sample-diamond interface, R$_{s-d}$($\omega$), at various pressures. Pressure evolution of the spectra are shown separately for the (a) low pressure (below 5.4 GPa) and (b) high pressure (above 5.4 GPa) regimes.}
\end{figure}

Figure 6 presents the mid-infrared reflectance spectra from the sample-diamond interface, $R_{s-d}(\omega)$, at various high pressures. Strong two-phonon diamond absorption prevents the measurement between 1800-2300 cm$^{-1}$. For the measurements at low pressures, sufficient initial pressure was applied to make the sample in good contact with diamond surface. The measured reflectance spectra for the polycrystalline sample shows the average behavior of that measured from different oriented single crystal sample~\cite{Moon2011}. On increasing pressure, the overall R$_{s-d}$ is found to enhance progressively. But the pressure variation of the reflectance spectra clearly shows two distinct regime. In the low pressure region (P$<$5.4 GPa), low frequency reflectance enhances rapidly compared to the high frequency part. Whereas, above 5.4 GPa, spectra in the high frequency region becomes highly pressure dependent. A sharp plasma edge at 2650 cm$^{-1}$ is observed in the reflectance spectra and this becomes more prominent upon increasing pressure.

\begin{figure}[tb]
\centerline{\includegraphics[width=80mm,clip]{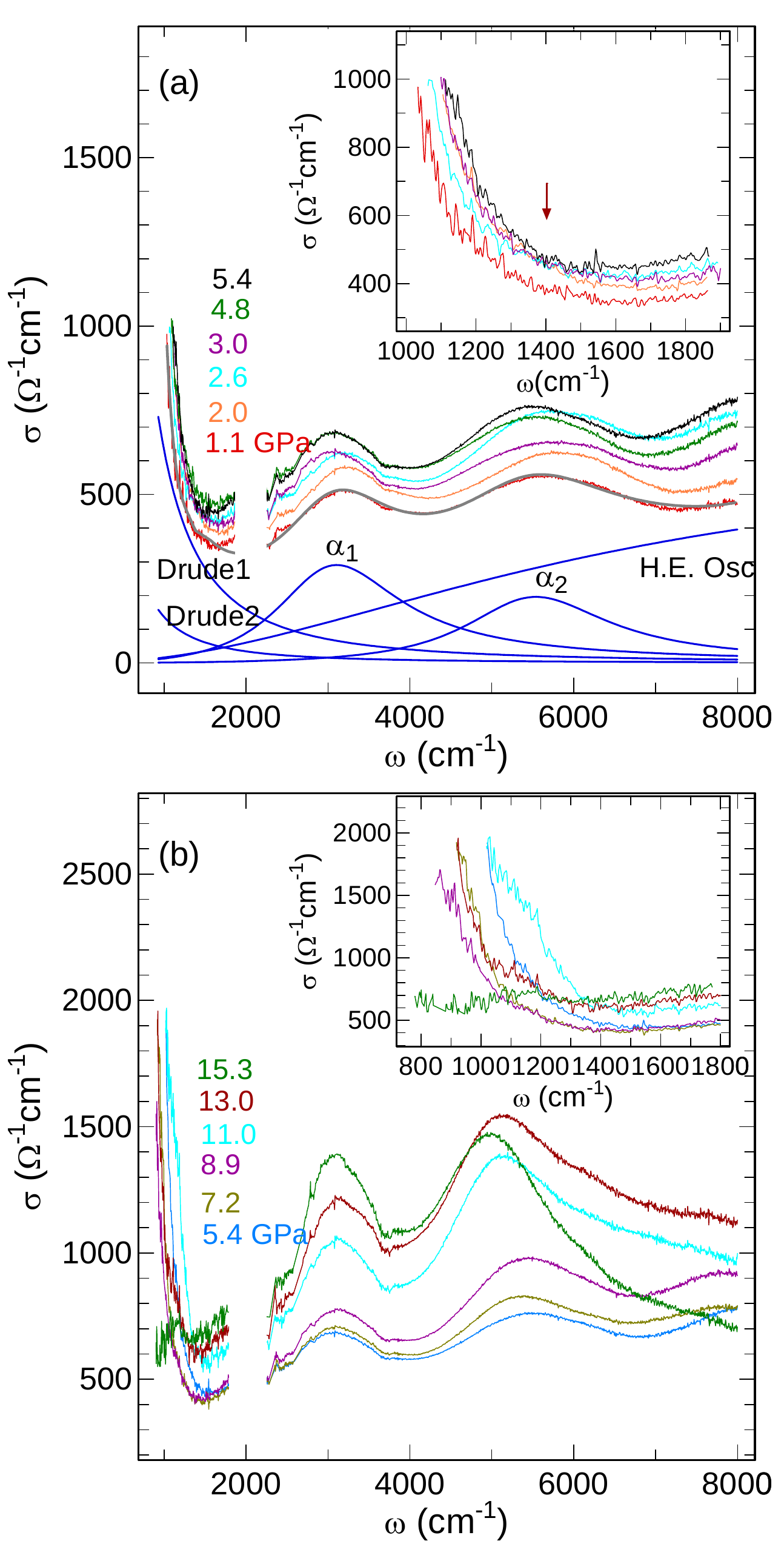}}
\caption{\label{fig5} (Color online) Frequency dependent optical conductivity (OC), $\sigma(\omega)$, at various pressures. (a) Pressure-variation of OC up to 5.4 GPa. A typical fit for the optical conductivity with superposition of Drude-Lorentz oscillators are shown for the lowest pressure data; Inset, anomalous change of the Drude spectral weight under pressure (b) Pressure-variation of OC above 5.4 GPa. Inset, dramatic change of Drude spectral weight across the structural transition. Pressure dependent color coding is maintained in the respective insets.}
\end{figure}

\begin{figure}[tb]
\centerline{\includegraphics[width=75mm,clip]{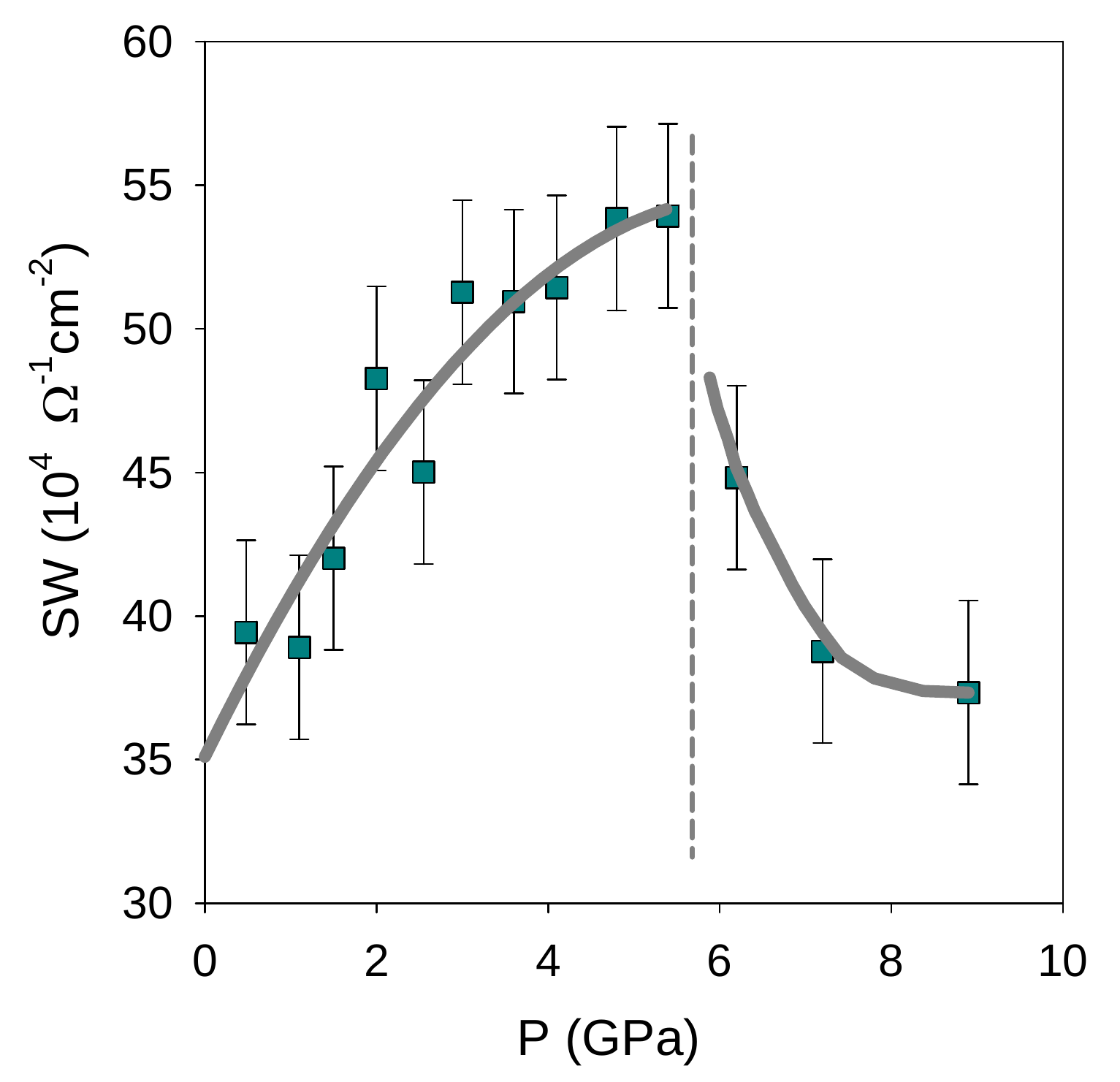}}
\caption{\label{fig5} (Color online) Pressure variation of the Drude spectral weight, $SW(P)=\int{\sigma(\omega,P)d\omega}$ integrated between 800 cm$^{-1}$ and 1800 cm$^{-1}$. The lines are only a guide to eye.}
\end{figure}

The optical conductivity (OC) spectra, $\sigma(\omega)$, plotted at various high pressures in figure 7, are calculated from $R_{s-d}(\omega)$ by using Kramer-Kronig transformation. For this purpose, the reflectance data is extrapolated to zero frequency based on a Drude-Lorentz fit and that to high frequency range with a high energy (at ~3 eV) Lorentz oscillator and at further higher frequency using the power law, $R(\omega)\propto\omega^{-4}$. Frequency dependent OC spectra are analyzed by fitting with superposition of Drude-Lorentz oscillators (by applying f -sum rule for auto-rescaling of the oscillator strength)~\cite{Kuzmany2009}. A typical fit is shown in Figure 7a for the low pressure OC spectra. Apart from the low frequency Drude free carrier response (composed of two terms, incoherent "Drude1" and coherent "Drude2"~\cite{Moon2010a,Wu2010,Mirri2012}) and a high energy broad mid-infrared band (shown by "H.E. Osc"), the OC comprises of two additional interband excitations ($\alpha1$, $\alpha2$) at 3100 cm$^{-1}$ and 5600 cm$^{-1}$, suggesting the multiband character of the system ~\cite{Boris2009,Mirri2012}. From density functional theory calculation, these bands are assigned to Fe 3d- Fe3d transitions~\cite{Moon2010a}. Comparing this spectra with that in ref. 35, it can be noted that with increased Te-doping, $\alpha1$ peak frequency decreases and oscillation strength increases, whereas $\alpha2$ peak frequency increases and its strength decreases. The over all $\sigma(\omega)$ is enhanced progressively upon increasing pressure up to 8.9 GPa. A dramatic change in the spectral feature is observed above this pressure. The high energy (mid-IR) band spectral weight reduces drastically and is transferred into the bands ($\alpha1$, $\alpha2$). Also these relatively narrow bands change their energy positions indicating a sudden change in the electronic structure of the system. These changes can be attributed to the tetragonal ($P4/nmm$) to orthorhombic ($Pbnm$) structural transition.


A closer look into the low frequency spectral feature in the OC shows anomalous changes at two pressures (2.6 GPa and 8.9 GPa), as shown in insets of Figure 7a,b. Although the overall spectral weight is enhanced systematically with pressure, the Drude spectral weight shows a pressure-induced crossover between 2.6 - 3 GPa around the isobestic (equal absorption) point at ~1400 cm$^{-1}$. Also the  This subtle change is due to the spectral rearrangement of the two Drude components by changing their spectral features (width, amplitude). A rapid electronic structural change is therefore speculated at this pressure. However, this subtle spectral change is not reflected in the pressure dependent total Drude spectral weight (SW), as can be seen in Figure 8. Due to the lack of low frequency reflectance data, the spectral feature of the individual Drude components cannot be obtained unambiguously. The total Drude spectral weight in the frequency range 800-1800 cm$^{-1}$, as determined by integrating the OC spectra, is taken as a measure of the effective free carrier density and its pressure evolution has been studied. Upon increasing pressure the total Drude spectral weight is found to increase monotonically up to 6 GPa. A slope change or a saturation of superconducting T$_{c}$ above 3 GPa may be explained in terms of the subtle electronic changes reflected in the low frequency optical conductivity at this pressure. Moreover the $\alpha1$, $\alpha2$ bands are also found to undergo a change from pressure dependent blue shift to red shift at this pressure. In inset of figure 7b, we clearly observe conspicuous changes in Drude spectral features above 5.4 GPa. There is a systematic reduction in the Drude spectral weight upon further increase in pressure (shown in figure 8). This is due to the emergence of the new phase having lower Drude spectral weight and systematic rise of its molar fraction. At 11 GPa, there is a sudden enhancement of the low frequency spectral weight due to the emergence of a strong absorption band of localized charges around 1150 cm$^{-1}$, which upon further increase of pressure gets suppressed and almost vanishes at 15 GPa. As this band appears above the structural transition pressure, this does not correspond to the high pressure orthorhombic phase. A link of this band emergence may be noticed with the sudden change of compressibility of the tetragonal phase and thus this feature may be related to the Fe$^{2+}$ spin-state transition.

Detailed structural information on the high pressure orthorhombic phase is given in ref. 9. Also, the high pressure investigation on undoped FeSe has indicated that the superconducting T$_{c}$ is enhanced in the high pressure $Pbnm$ phase~\cite{Garbarino2009}. It may therefore be speculated that the relative phase fraction of the high pressure orthorhombic phase is also responsible for the discrepancies in the pressure dependencies of T$_{c}$ in various studies of FeSe$_{0.5}$Te$_{0.5}$, apart from the composition stoichiometry variation and pressure inaccuracy~\cite{Gresty2009,Huang2009,Gooch2012,Tsoi2009,Karmakar2013}. Our results suggest that although the orthorhombic phase starts emerging at $\sim$4 GPa, the phase fraction becomes significant only at $\sim$6 GPa where the over all superconducting feature is taken over by the this new phase. A mixed phase behaviour is further supported by the observed broadening of the superconducting transition. A detailed band structure analysis of the high pressure orthorhombic phase and its comparison with the tetragonal phase~\cite{Subedi2008,Charnukha2011} would therefore shed light in order to further clarify the pressure evolution of the superconductivity in the iron-chalcogenides.\\

In conclusion, we have found that the superconductivity onset in tetragonal Fe$_{1.03}$Se$_{0.5}$Te$_{0.5}$ attains a broad maximum value of 22.4K at $\sim$4 GPa. The high pressure orthorhombic phase (S.G. $Pbnm$, z=4) starts emerging at a low pressure ($\sim$4 GPa), but the transition becomes apparent only above 10 GPa. The tetragonal phase is found to be extremely soft and the pressure response of the structure reveals a strong correlation between the Fe(Se,Te)$_{4}$ tetrahedral deformation and the superconducting T$_{c}$. Subtle structural modification and change in bulk compressibility of the existing tetragonal phase above 10 GPa suggests a new tetragonal structure with possible Fe$^{2+}$ spin-state transition at high pressure. Pressure evolution of optical conductivity clearly indicates dramatic change in electronic structure of the system across the structural transition. An anomalous and subtle change in low frequency Drude spectral weight at $\sim$3 GPa indicates a sudden rearrangement of the electronic band structure of the tetragonal phase causing the saturation of T$_{c}$ above this pressure. Detailed band structure calculation of the orthorhombic $\textit{Pbnm}$ phase is prescribed to understand the magnetism and superconductivity of the high pressure phase of iron-chalcogenide compounds.\

\begin{acknowledgments}
We acknowledge the Elettra Synchrotron facility for the provision of the beamtime. We like to acknowledge H. Bhatt, A.K. Mishra and Debjani Karmakar (BARC), A. Peruchhi and P. Maurizio (Elettra) for their help during high pressure measurements. Financial supports by Department of Science and Technology (Government of India) and ICTP (Italy) is also gratefully acknowledged.

\end{acknowledgments}

\bibliography{fesete}

\begin{thebibliography}{40}
\expandafter\ifx\csname natexlab\endcsname\relax\def\natexlab#1{#1}\fi
\expandafter\ifx\csname bibnamefont\endcsname\relax
  \def\bibnamefont#1{#1}\fi
\expandafter\ifx\csname bibfnamefont\endcsname\relax
  \def\bibfnamefont#1{#1}\fi
\expandafter\ifx\csname citenamefont\endcsname\relax
  \def\citenamefont#1{#1}\fi
\expandafter\ifx\csname url\endcsname\relax
  \def\url#1{\texttt{#1}}\fi
\expandafter\ifx\csname urlprefix\endcsname\relax\def\urlprefix{URL }\fi
\providecommand{\bibinfo}[2]{#2}
\providecommand{\eprint}[2][]{\url{#2}}

\bibitem[{\citenamefont{Hsu et~al.}(2008)\citenamefont{Hsu, Luo, Yeh, Chen,
  Huang, Wu, Lee, Huang, Chu, Yan et~al.}}]{Hsu2008}
\bibinfo{author}{\bibfnamefont{F.-C.} \bibnamefont{Hsu}},
  \bibinfo{author}{\bibfnamefont{J.-Y.} \bibnamefont{Luo}},
  \bibinfo{author}{\bibfnamefont{K.-W.} \bibnamefont{Yeh}},
  \bibinfo{author}{\bibfnamefont{T.-K.} \bibnamefont{Chen}},
  \bibinfo{author}{\bibfnamefont{T.-W.} \bibnamefont{Huang}},
  \bibinfo{author}{\bibfnamefont{P.~M.} \bibnamefont{Wu}},
  \bibinfo{author}{\bibfnamefont{Y.-C.} \bibnamefont{Lee}},
  \bibinfo{author}{\bibfnamefont{Y.-L.} \bibnamefont{Huang}},
  \bibinfo{author}{\bibfnamefont{Y.-Y.} \bibnamefont{Chu}},
  \bibinfo{author}{\bibfnamefont{D.-C.} \bibnamefont{Yan}},
  \bibnamefont{et~al.}, \bibinfo{journal}{Proc. Natl. Acad. Sci.}
  \textbf{\bibinfo{volume}{105}}, \bibinfo{pages}{14262}
  (\bibinfo{year}{2008}).

\bibitem[{\citenamefont{Imai et~al.}(2009)\citenamefont{Imai, Ahilan, Ning,
  McQueen, and Cava}}]{Imai2009}
\bibinfo{author}{\bibfnamefont{T.}~\bibnamefont{Imai}},
  \bibinfo{author}{\bibfnamefont{K.}~\bibnamefont{Ahilan}},
  \bibinfo{author}{\bibfnamefont{F.~L.} \bibnamefont{Ning}},
  \bibinfo{author}{\bibfnamefont{T.~M.} \bibnamefont{McQueen}},
  \bibnamefont{and} \bibinfo{author}{\bibfnamefont{R.~J.} \bibnamefont{Cava}},
  \bibinfo{journal}{Phys. Rev. Lett.} \textbf{\bibinfo{volume}{102}},
  \bibinfo{pages}{177005} (\bibinfo{year}{2009}).

\bibitem[{\citenamefont{Bendele et~al.}(2010)\citenamefont{Bendele, Amato,
  Conder, Elender, Keller, Klauss, Luetkens, Pomjakushina, Raselli, and
  Khasanov}}]{Bendele2010}
\bibinfo{author}{\bibfnamefont{M.}~\bibnamefont{Bendele}},
  \bibinfo{author}{\bibfnamefont{A.}~\bibnamefont{Amato}},
  \bibinfo{author}{\bibfnamefont{K.}~\bibnamefont{Conder}},
  \bibinfo{author}{\bibfnamefont{M.}~\bibnamefont{Elender}},
  \bibinfo{author}{\bibfnamefont{H.}~\bibnamefont{Keller}},
  \bibinfo{author}{\bibfnamefont{H.-H.} \bibnamefont{Klauss}},
  \bibinfo{author}{\bibfnamefont{H.}~\bibnamefont{Luetkens}},
  \bibinfo{author}{\bibfnamefont{E.}~\bibnamefont{Pomjakushina}},
  \bibinfo{author}{\bibfnamefont{A.}~\bibnamefont{Raselli}}, \bibnamefont{and}
  \bibinfo{author}{\bibfnamefont{R.}~\bibnamefont{Khasanov}},
  \bibinfo{journal}{Phys. Rev. Lett.} \textbf{\bibinfo{volume}{104}},
  \bibinfo{pages}{087003} (\bibinfo{year}{2010}).

\bibitem[{\citenamefont{Mizuguchi et~al.}(2008)\citenamefont{Mizuguchi,
  Tomioka, Tsuda, Yamaguchi, and Takano}}]{Mizuguchi2008}
\bibinfo{author}{\bibfnamefont{Y.}~\bibnamefont{Mizuguchi}},
  \bibinfo{author}{\bibfnamefont{F.}~\bibnamefont{Tomioka}},
  \bibinfo{author}{\bibfnamefont{S.}~\bibnamefont{Tsuda}},
  \bibinfo{author}{\bibfnamefont{T.}~\bibnamefont{Yamaguchi}},
  \bibnamefont{and} \bibinfo{author}{\bibfnamefont{Y.}~\bibnamefont{Takano}},
  \bibinfo{journal}{Appl. Phys. Lett.} \textbf{\bibinfo{volume}{93}},
  \bibinfo{pages}{152505} (\bibinfo{year}{2008}).

\bibitem[{\citenamefont{Miyoshi et~al.}(2009)\citenamefont{Miyoshi, Takaichi,
  Mutou, Fujiwara, and Takeuchi}}]{Miyoshi2009}
\bibinfo{author}{\bibfnamefont{K.}~\bibnamefont{Miyoshi}},
  \bibinfo{author}{\bibfnamefont{Y.}~\bibnamefont{Takaichi}},
  \bibinfo{author}{\bibfnamefont{E.}~\bibnamefont{Mutou}},
  \bibinfo{author}{\bibfnamefont{K.}~\bibnamefont{Fujiwara}}, \bibnamefont{and}
  \bibinfo{author}{\bibfnamefont{J.}~\bibnamefont{Takeuchi}},
  \bibinfo{journal}{J. Phys. Soc. Jpn.} \textbf{\bibinfo{volume}{78}},
  \bibinfo{pages}{093703} (\bibinfo{year}{2009}).

\bibitem[{\citenamefont{Huang et~al.}(2009)\citenamefont{Huang, Chou, Tseng,
  Huang, Hsu, Yeh, Wu, and Yang}}]{Huang2009}
\bibinfo{author}{\bibfnamefont{C.}~\bibnamefont{Huang}},
  \bibinfo{author}{\bibfnamefont{C.-C.} \bibnamefont{Chou}},
  \bibinfo{author}{\bibfnamefont{K.-F.} \bibnamefont{Tseng}},
  \bibinfo{author}{\bibfnamefont{Y.-L.} \bibnamefont{Huang}},
  \bibinfo{author}{\bibfnamefont{F.-C.} \bibnamefont{Hsu}},
  \bibinfo{author}{\bibfnamefont{K.-W.} \bibnamefont{Yeh}},
  \bibinfo{author}{\bibfnamefont{M.-K.} \bibnamefont{Wu}}, \bibnamefont{and}
  \bibinfo{author}{\bibfnamefont{H.-D.} \bibnamefont{Yang}},
  \bibinfo{journal}{J. Phys. Soc. Jpn.} \textbf{\bibinfo{volume}{78}},
  \bibinfo{pages}{084710} (\bibinfo{year}{2009}).

\bibitem[{\citenamefont{Okabe et~al.}(2010)\citenamefont{Okabe, Takeshita,
  Horigane, Muranaka, and Akimitsu}}]{Okabe2010}
\bibinfo{author}{\bibfnamefont{H.}~\bibnamefont{Okabe}},
  \bibinfo{author}{\bibfnamefont{N.}~\bibnamefont{Takeshita}},
  \bibinfo{author}{\bibfnamefont{K.}~\bibnamefont{Horigane}},
  \bibinfo{author}{\bibfnamefont{T.}~\bibnamefont{Muranaka}}, \bibnamefont{and}
  \bibinfo{author}{\bibfnamefont{J.}~\bibnamefont{Akimitsu}},
  \bibinfo{journal}{Phys. Rev. B} \textbf{\bibinfo{volume}{81}},
  \bibinfo{pages}{205119} (\bibinfo{year}{2010}).

\bibitem[{\citenamefont{Uhoya et~al.}(2012)\citenamefont{Uhoya, Tsoi, Vohra,
  Wolanyk, Rao, Wu, and Weir}}]{Uhoya2012}
\bibinfo{author}{\bibfnamefont{W.}~\bibnamefont{Uhoya}},
  \bibinfo{author}{\bibfnamefont{G.}~\bibnamefont{Tsoi}},
  \bibinfo{author}{\bibfnamefont{Y.}~\bibnamefont{Vohra}},
  \bibinfo{author}{\bibfnamefont{N.}~\bibnamefont{Wolanyk}},
  \bibinfo{author}{\bibfnamefont{S.~M.} \bibnamefont{Rao}},
  \bibinfo{author}{\bibfnamefont{M.-K.} \bibnamefont{Wu}}, \bibnamefont{and}
  \bibinfo{author}{\bibfnamefont{S.}~\bibnamefont{Weir}},
  \bibinfo{journal}{Euro. Phys. Lett.} \textbf{\bibinfo{volume}{99}},
  \bibinfo{pages}{26002} (\bibinfo{year}{2012}).

\bibitem[{\citenamefont{Kumar et~al.}(2010)\citenamefont{Kumar, Zhang,
  Sinogeiken, Xiao, Kumar, Chow, Cornelius, and Chen}}]{Kumar2010}
\bibinfo{author}{\bibfnamefont{R.~S.} \bibnamefont{Kumar}},
  \bibinfo{author}{\bibfnamefont{Y.}~\bibnamefont{Zhang}},
  \bibinfo{author}{\bibfnamefont{S.}~\bibnamefont{Sinogeiken}},
  \bibinfo{author}{\bibfnamefont{Y.}~\bibnamefont{Xiao}},
  \bibinfo{author}{\bibfnamefont{S.}~\bibnamefont{Kumar}},
  \bibinfo{author}{\bibfnamefont{P.}~\bibnamefont{Chow}},
  \bibinfo{author}{\bibfnamefont{A.~L.} \bibnamefont{Cornelius}},
  \bibnamefont{and} \bibinfo{author}{\bibfnamefont{C.}~\bibnamefont{Chen}},
  \bibinfo{journal}{J. Phys. Chem. B} \textbf{\bibinfo{volume}{114}},
  \bibinfo{pages}{12597} (\bibinfo{year}{2010}).

\bibitem[{\citenamefont{Garbarino et~al.}(2009)\citenamefont{Garbarino, Sow,
  Lejay, Sulpice, Toulemonde, Mezouar, and Núñez-Regueiro}}]{Garbarino2009}
\bibinfo{author}{\bibfnamefont{G.}~\bibnamefont{Garbarino}},
  \bibinfo{author}{\bibfnamefont{A.}~\bibnamefont{Sow}},
  \bibinfo{author}{\bibfnamefont{P.}~\bibnamefont{Lejay}},
  \bibinfo{author}{\bibfnamefont{A.}~\bibnamefont{Sulpice}},
  \bibinfo{author}{\bibfnamefont{P.}~\bibnamefont{Toulemonde}},
  \bibinfo{author}{\bibfnamefont{M.}~\bibnamefont{Mezouar}}, \bibnamefont{and}
  \bibinfo{author}{\bibfnamefont{M.}~\bibnamefont{Núñez-Regueiro}},
  \bibinfo{journal}{Euro. Phys. Lett.} \textbf{\bibinfo{volume}{86}},
  \bibinfo{pages}{27001} (\bibinfo{year}{2009}).

\bibitem[{\citenamefont{Braithwaite et~al.}(2009)\citenamefont{Braithwaite,
  Salce, Lapertot, Bourdarot, Marin, Aoki, and Hanfland}}]{Braithwaite2009}
\bibinfo{author}{\bibfnamefont{D.}~\bibnamefont{Braithwaite}},
  \bibinfo{author}{\bibfnamefont{B.}~\bibnamefont{Salce}},
  \bibinfo{author}{\bibfnamefont{G.}~\bibnamefont{Lapertot}},
  \bibinfo{author}{\bibfnamefont{F.}~\bibnamefont{Bourdarot}},
  \bibinfo{author}{\bibfnamefont{C.}~\bibnamefont{Marin}},
  \bibinfo{author}{\bibfnamefont{D.}~\bibnamefont{Aoki}}, \bibnamefont{and}
  \bibinfo{author}{\bibfnamefont{M.}~\bibnamefont{Hanfland}},
  \bibinfo{journal}{J. Phys.: Condens. Matter} \textbf{\bibinfo{volume}{21}},
  \bibinfo{pages}{232202} (\bibinfo{year}{2009}).

\bibitem[{\citenamefont{Margadonna et~al.}(2009)\citenamefont{Margadonna,
  Takabayashi, Ohishi, Mizuguchi, Takano, Kagayama, Nakagawa, Takata, and
  Prassides}}]{Margadonna2009}
\bibinfo{author}{\bibfnamefont{S.}~\bibnamefont{Margadonna}},
  \bibinfo{author}{\bibfnamefont{Y.}~\bibnamefont{Takabayashi}},
  \bibinfo{author}{\bibfnamefont{Y.}~\bibnamefont{Ohishi}},
  \bibinfo{author}{\bibfnamefont{Y.}~\bibnamefont{Mizuguchi}},
  \bibinfo{author}{\bibfnamefont{Y.}~\bibnamefont{Takano}},
  \bibinfo{author}{\bibfnamefont{T.}~\bibnamefont{Kagayama}},
  \bibinfo{author}{\bibfnamefont{T.}~\bibnamefont{Nakagawa}},
  \bibinfo{author}{\bibfnamefont{M.}~\bibnamefont{Takata}}, \bibnamefont{and}
  \bibinfo{author}{\bibfnamefont{K.}~\bibnamefont{Prassides}},
  \bibinfo{journal}{Phys. Rev. B} \textbf{\bibinfo{volume}{80}},
  \bibinfo{pages}{064506} (\bibinfo{year}{2009}).

\bibitem[{\citenamefont{Medvedev et~al.}(2009)\citenamefont{Medvedev, McQueen,
  Troyan, Palasyuk, Eremets, Cava, Naghavi, Casper, Ksenofontov, G.Wortmann
  et~al.}}]{Medvedev2009}
\bibinfo{author}{\bibfnamefont{S.}~\bibnamefont{Medvedev}},
  \bibinfo{author}{\bibfnamefont{T.~M.} \bibnamefont{McQueen}},
  \bibinfo{author}{\bibfnamefont{I.~A.} \bibnamefont{Troyan}},
  \bibinfo{author}{\bibfnamefont{T.}~\bibnamefont{Palasyuk}},
  \bibinfo{author}{\bibfnamefont{M.~I.} \bibnamefont{Eremets}},
  \bibinfo{author}{\bibfnamefont{R.~J.} \bibnamefont{Cava}},
  \bibinfo{author}{\bibfnamefont{S.}~\bibnamefont{Naghavi}},
  \bibinfo{author}{\bibfnamefont{F.}~\bibnamefont{Casper}},
  \bibinfo{author}{\bibfnamefont{V.}~\bibnamefont{Ksenofontov}},
  \bibinfo{author}{\bibnamefont{G.Wortmann}}, \bibnamefont{et~al.},
  \bibinfo{journal}{Nature Materials} \textbf{\bibinfo{volume}{8}},
  \bibinfo{pages}{630} (\bibinfo{year}{2009}).

\bibitem[{\citenamefont{Stemshorn et~al.}(2010)\citenamefont{Stemshorn, Tsoi,
  Vohra, Sainogeiken, Wu, Huang, Rao, Wu, Yeh, and Weir}}]{Stemshorn2010}
\bibinfo{author}{\bibfnamefont{A.~K.} \bibnamefont{Stemshorn}},
  \bibinfo{author}{\bibfnamefont{G.}~\bibnamefont{Tsoi}},
  \bibinfo{author}{\bibfnamefont{Y.~K.} \bibnamefont{Vohra}},
  \bibinfo{author}{\bibfnamefont{S.}~\bibnamefont{Sainogeiken}},
  \bibinfo{author}{\bibfnamefont{P.~M.} \bibnamefont{Wu}},
  \bibinfo{author}{\bibfnamefont{Y.}~\bibnamefont{Huang}},
  \bibinfo{author}{\bibfnamefont{S.~M.} \bibnamefont{Rao}},
  \bibinfo{author}{\bibfnamefont{M.-K.} \bibnamefont{Wu}},
  \bibinfo{author}{\bibfnamefont{K.~W.} \bibnamefont{Yeh}}, \bibnamefont{and}
  \bibinfo{author}{\bibfnamefont{S.~T.} \bibnamefont{Weir}},
  \bibinfo{journal}{J. Mater. Res.} \textbf{\bibinfo{volume}{25}},
  \bibinfo{pages}{396} (\bibinfo{year}{2010}).

\bibitem[{\citenamefont{Yeh et~al.}(2008)\citenamefont{Yeh, Huang, l.~Huang,
  Chen, Hsu, Wu, Lee, Chu, Chen, Luo et~al.}}]{Yeh2008}
\bibinfo{author}{\bibfnamefont{K.-W.} \bibnamefont{Yeh}},
  \bibinfo{author}{\bibfnamefont{T.-W.} \bibnamefont{Huang}},
  \bibinfo{author}{\bibfnamefont{Y.}~\bibnamefont{l.~Huang}},
  \bibinfo{author}{\bibfnamefont{T.-K.} \bibnamefont{Chen}},
  \bibinfo{author}{\bibfnamefont{F.-C.} \bibnamefont{Hsu}},
  \bibinfo{author}{\bibfnamefont{P.~M.} \bibnamefont{Wu}},
  \bibinfo{author}{\bibfnamefont{Y.-C.} \bibnamefont{Lee}},
  \bibinfo{author}{\bibfnamefont{Y.-Y.} \bibnamefont{Chu}},
  \bibinfo{author}{\bibfnamefont{C.-L.} \bibnamefont{Chen}},
  \bibinfo{author}{\bibfnamefont{J.-Y.} \bibnamefont{Luo}},
  \bibnamefont{et~al.}, \bibinfo{journal}{Euro. Phys. Lett.}
  \textbf{\bibinfo{volume}{84}}, \bibinfo{pages}{37002} (\bibinfo{year}{2008}).

\bibitem[{\citenamefont{Horigane et~al.}(2009)\citenamefont{Horigane, Hiraka,
  and Ohoyama}}]{Horigane2009}
\bibinfo{author}{\bibfnamefont{K.}~\bibnamefont{Horigane}},
  \bibinfo{author}{\bibfnamefont{H.}~\bibnamefont{Hiraka}}, \bibnamefont{and}
  \bibinfo{author}{\bibfnamefont{K.}~\bibnamefont{Ohoyama}},
  \bibinfo{journal}{J. Phys. Soc. Jpn.} \textbf{\bibinfo{volume}{78}},
  \bibinfo{pages}{074718} (\bibinfo{year}{2009}).

\bibitem[{\citenamefont{Gresty et~al.}(2009)\citenamefont{Gresty, Takabayashi,
  Ganin, McDonald, Claridge, Giap, Mizuguchi, Takano, Kagayama, Ohishi
  et~al.}}]{Gresty2009}
\bibinfo{author}{\bibfnamefont{N.~C.} \bibnamefont{Gresty}},
  \bibinfo{author}{\bibfnamefont{Y.}~\bibnamefont{Takabayashi}},
  \bibinfo{author}{\bibfnamefont{A.~Y.} \bibnamefont{Ganin}},
  \bibinfo{author}{\bibfnamefont{M.~T.} \bibnamefont{McDonald}},
  \bibinfo{author}{\bibfnamefont{J.~B.} \bibnamefont{Claridge}},
  \bibinfo{author}{\bibfnamefont{D.}~\bibnamefont{Giap}},
  \bibinfo{author}{\bibfnamefont{Y.}~\bibnamefont{Mizuguchi}},
  \bibinfo{author}{\bibfnamefont{Y.}~\bibnamefont{Takano}},
  \bibinfo{author}{\bibfnamefont{T.}~\bibnamefont{Kagayama}},
  \bibinfo{author}{\bibfnamefont{Y.}~\bibnamefont{Ohishi}},
  \bibnamefont{et~al.}, \bibinfo{journal}{J. Am. Chem. Soc.}
  \textbf{\bibinfo{volume}{131}}, \bibinfo{pages}{16944}
  (\bibinfo{year}{2009}).

\bibitem[{\citenamefont{Gooch et~al.}(2012)\citenamefont{Gooch, Lorenz, Huang,
  Chien, and Chu}}]{Gooch2012}
\bibinfo{author}{\bibfnamefont{M.}~\bibnamefont{Gooch}},
  \bibinfo{author}{\bibfnamefont{B.}~\bibnamefont{Lorenz}},
  \bibinfo{author}{\bibfnamefont{S.~X.} \bibnamefont{Huang}},
  \bibinfo{author}{\bibfnamefont{C.~L.} \bibnamefont{Chien}}, \bibnamefont{and}
  \bibinfo{author}{\bibfnamefont{C.~W.} \bibnamefont{Chu}},
  \bibinfo{journal}{J. Appl. Phys.} \textbf{\bibinfo{volume}{111}},
  \bibinfo{pages}{112610} (\bibinfo{year}{2012}).

\bibitem[{\citenamefont{Tsoi et~al.}(2009)\citenamefont{Tsoi, Stemshorn, Vohra,
  Wu, Hsu, Huang, Wu, Yeh, and Weir}}]{Tsoi2009}
\bibinfo{author}{\bibfnamefont{G.}~\bibnamefont{Tsoi}},
  \bibinfo{author}{\bibfnamefont{A.~K.} \bibnamefont{Stemshorn}},
  \bibinfo{author}{\bibfnamefont{Y.~K.} \bibnamefont{Vohra}},
  \bibinfo{author}{\bibfnamefont{P.~M.} \bibnamefont{Wu}},
  \bibinfo{author}{\bibfnamefont{F.~C.} \bibnamefont{Hsu}},
  \bibinfo{author}{\bibfnamefont{Y.~L.} \bibnamefont{Huang}},
  \bibinfo{author}{\bibfnamefont{M.~K.} \bibnamefont{Wu}},
  \bibinfo{author}{\bibfnamefont{K.~W.} \bibnamefont{Yeh}}, \bibnamefont{and}
  \bibinfo{author}{\bibfnamefont{S.~T.} \bibnamefont{Weir}},
  \bibinfo{journal}{J. Phys.: Condens. Matter} \textbf{\bibinfo{volume}{21}},
  \bibinfo{pages}{232201} (\bibinfo{year}{2009}).

\bibitem[{\citenamefont{Karmakar}(2013)}]{Karmakar2013}
\bibinfo{author}{\bibfnamefont{S.}~\bibnamefont{Karmakar}},
  \bibinfo{journal}{High Pressure Research} \textbf{\bibinfo{volume}{33}},
  \bibinfo{pages}{381} (\bibinfo{year}{2013}).

\bibitem[{\citenamefont{Stemshorn et~al.}(2009)\citenamefont{Stemshorn, Vohra,
  Wu, Hsu, Huang, Wu, and Yeh}}]{Stemshorn2009}
\bibinfo{author}{\bibfnamefont{A.~K.} \bibnamefont{Stemshorn}},
  \bibinfo{author}{\bibfnamefont{Y.~K.} \bibnamefont{Vohra}},
  \bibinfo{author}{\bibfnamefont{P.~M.} \bibnamefont{Wu}},
  \bibinfo{author}{\bibfnamefont{F.~C.} \bibnamefont{Hsu}},
  \bibinfo{author}{\bibfnamefont{Y.~L.} \bibnamefont{Huang}},
  \bibinfo{author}{\bibfnamefont{M.~K.} \bibnamefont{Wu}}, \bibnamefont{and}
  \bibinfo{author}{\bibfnamefont{K.~W.} \bibnamefont{Yeh}},
  \bibinfo{journal}{High Pressure Research} \textbf{\bibinfo{volume}{29}},
  \bibinfo{pages}{267} (\bibinfo{year}{2009}).

\bibitem[{\citenamefont{Moon and Choi}(2010)}]{Moon2010}
\bibinfo{author}{\bibfnamefont{C.-Y.} \bibnamefont{Moon}} \bibnamefont{and}
  \bibinfo{author}{\bibfnamefont{H.~J.} \bibnamefont{Choi}},
  \bibinfo{journal}{Phys. Rev. Lett.} \textbf{\bibinfo{volume}{104}},
  \bibinfo{pages}{057003} (\bibinfo{year}{2010}).

\bibitem[{\citenamefont{Kumar et~al.}(2012)\citenamefont{Kumar, Auluck,
  Ahluwalia, and Awana}}]{Kumar2012}
\bibinfo{author}{\bibfnamefont{J.}~\bibnamefont{Kumar}},
  \bibinfo{author}{\bibfnamefont{S.}~\bibnamefont{Auluck}},
  \bibinfo{author}{\bibfnamefont{P.~K.} \bibnamefont{Ahluwalia}},
  \bibnamefont{and} \bibinfo{author}{\bibfnamefont{V.~P.~S.}
  \bibnamefont{Awana}}, \bibinfo{journal}{Supercond. Sci. Technol.}
  \textbf{\bibinfo{volume}{25}}, \bibinfo{pages}{095002}
  (\bibinfo{year}{2012}).

\bibitem[{\citenamefont{Ciechan et~al.}(2012)\citenamefont{Ciechan, Winiarski,
  and Samsel-Czeka}}]{Ciechan2012}
\bibinfo{author}{\bibfnamefont{A.}~\bibnamefont{Ciechan}},
  \bibinfo{author}{\bibfnamefont{M.}~\bibnamefont{Winiarski}},
  \bibnamefont{and}
  \bibinfo{author}{\bibfnamefont{M.}~\bibnamefont{Samsel-Czeka}},
  \bibinfo{journal}{Acta Physica Polonica A} \textbf{\bibinfo{volume}{121}},
  \bibinfo{pages}{820} (\bibinfo{year}{2012}).

\bibitem[{\citenamefont{Dewaele et~al.}(2004)\citenamefont{Dewaele, Loubeyre,
  and Mezouar}}]{Dewaele2004}
\bibinfo{author}{\bibfnamefont{A.}~\bibnamefont{Dewaele}},
  \bibinfo{author}{\bibfnamefont{P.}~\bibnamefont{Loubeyre}}, \bibnamefont{and}
  \bibinfo{author}{\bibfnamefont{M.}~\bibnamefont{Mezouar}},
  \bibinfo{journal}{Phys. Rev. B} \textbf{\bibinfo{volume}{70}},
  \bibinfo{pages}{094112} (\bibinfo{year}{2004}).

\bibitem[{\citenamefont{Hammersley}(1998)}]{Hammersley1998}
\bibinfo{author}{\bibfnamefont{A.}~\bibnamefont{Hammersley}},
  \emph{\bibinfo{title}{Computer Program FIT2D}} (\bibinfo{publisher}{ESRF,
  Grenoble}, \bibinfo{year}{1998}).

\bibitem[{\citenamefont{Larson and von Dreele}(1986)}]{Larson1986}
\bibinfo{author}{\bibfnamefont{A.~C.} \bibnamefont{Larson}} \bibnamefont{and}
  \bibinfo{author}{\bibfnamefont{R.~B.} \bibnamefont{von Dreele}},
  \emph{\bibinfo{title}{Report No. LAUR 86-748}} (\bibinfo{publisher}{Los
  Alamos National Lab, New Mexico}, \bibinfo{year}{1986}).

\bibitem[{\citenamefont{Mao et~al.}(1986)\citenamefont{Mao, Xu, and
  Bell}}]{Mao1986}
\bibinfo{author}{\bibfnamefont{H.}~\bibnamefont{Mao}},
  \bibinfo{author}{\bibfnamefont{J.}~\bibnamefont{Xu}}, \bibnamefont{and}
  \bibinfo{author}{\bibfnamefont{P.}~\bibnamefont{Bell}}, \bibinfo{journal}{J.
  Geophys. Res.} \textbf{\bibinfo{volume}{91}}, \bibinfo{pages}{4673}
  (\bibinfo{year}{1986}).

\bibitem[{\citenamefont{Birch}(1978)}]{Birch1978}
\bibinfo{author}{\bibfnamefont{F.}~\bibnamefont{Birch}}, \bibinfo{journal}{J.
  Geophys. Res.} \textbf{\bibinfo{volume}{83}}, \bibinfo{pages}{1257}
  (\bibinfo{year}{1978}).

\bibitem[{\citenamefont{Kumar et~al.}(2011)\citenamefont{Kumar, Zhang, Xiao,
  Baker, Cornelius, Veeramalai, Chow, Chen, and Zhao}}]{Kumar2011}
\bibinfo{author}{\bibfnamefont{R.~S.} \bibnamefont{Kumar}},
  \bibinfo{author}{\bibfnamefont{Y.}~\bibnamefont{Zhang}},
  \bibinfo{author}{\bibfnamefont{Y.}~\bibnamefont{Xiao}},
  \bibinfo{author}{\bibfnamefont{J.}~\bibnamefont{Baker}},
  \bibinfo{author}{\bibfnamefont{A.}~\bibnamefont{Cornelius}},
  \bibinfo{author}{\bibfnamefont{S.}~\bibnamefont{Veeramalai}},
  \bibinfo{author}{\bibfnamefont{P.}~\bibnamefont{Chow}},
  \bibinfo{author}{\bibfnamefont{C.}~\bibnamefont{Chen}}, \bibnamefont{and}
  \bibinfo{author}{\bibfnamefont{Y.}~\bibnamefont{Zhao}},
  \bibinfo{journal}{Appl. Phys. Lett.} \textbf{\bibinfo{volume}{99}},
  \bibinfo{pages}{061913} (\bibinfo{year}{2011}).

\bibitem[{\citenamefont{Tamai et~al.}(2010)\citenamefont{Tamai, Ganin,
  Rozbicki, Bacsa, Meevasana, King, Caffio, Schaub, and
  Margadonna}}]{Tamai2010}
\bibinfo{author}{\bibfnamefont{A.}~\bibnamefont{Tamai}},
  \bibinfo{author}{\bibfnamefont{A.}~\bibnamefont{Ganin}},
  \bibinfo{author}{\bibfnamefont{E.}~\bibnamefont{Rozbicki}},
  \bibinfo{author}{\bibfnamefont{J.}~\bibnamefont{Bacsa}},
  \bibinfo{author}{\bibfnamefont{W.}~\bibnamefont{Meevasana}},
  \bibinfo{author}{\bibfnamefont{P.~D.~C.} \bibnamefont{King}},
  \bibinfo{author}{\bibfnamefont{M.}~\bibnamefont{Caffio}},
  \bibinfo{author}{\bibfnamefont{R.}~\bibnamefont{Schaub}}, \bibnamefont{and}
  \bibinfo{author}{\bibfnamefont{S.}~\bibnamefont{Margadonna}},
  \bibinfo{journal}{Phys. Rev. Lett.} \textbf{\bibinfo{volume}{104}},
  \bibinfo{pages}{097002} (\bibinfo{year}{2010}).

\bibitem[{\citenamefont{Schilling}(2007)}]{Schilling2006}
\bibinfo{author}{\bibfnamefont{J.~S.} \bibnamefont{Schilling}},
  \emph{\bibinfo{title}{High Pressure Effects in "High-Temperature
  Superconductivity: A Treatise on Theory and Applications", Ed. J. R.
  Schrieffer and J. S. Brooks, p 427}} (\bibinfo{publisher}{Springer, Berlin},
  \bibinfo{year}{2007}).

\bibitem[{\citenamefont{Moon et~al.}(2011)\citenamefont{Moon, Homes, Akrap, Xu,
  Wen, Lin, Li, Gu, and Basov}}]{Moon2011}
\bibinfo{author}{\bibfnamefont{S.~J.} \bibnamefont{Moon}},
  \bibinfo{author}{\bibfnamefont{C.~C.} \bibnamefont{Homes}},
  \bibinfo{author}{\bibfnamefont{A.}~\bibnamefont{Akrap}},
  \bibinfo{author}{\bibfnamefont{Z.~J.} \bibnamefont{Xu}},
  \bibinfo{author}{\bibfnamefont{J.~S.} \bibnamefont{Wen}},
  \bibinfo{author}{\bibfnamefont{Z.~W.} \bibnamefont{Lin}},
  \bibinfo{author}{\bibfnamefont{Q.}~\bibnamefont{Li}},
  \bibinfo{author}{\bibfnamefont{G.~D.} \bibnamefont{Gu}}, \bibnamefont{and}
  \bibinfo{author}{\bibfnamefont{D.~N.} \bibnamefont{Basov}},
  \bibinfo{journal}{Phys. Rev. Lett.} \textbf{\bibinfo{volume}{106}},
  \bibinfo{pages}{217001} (\bibinfo{year}{2011}).

\bibitem[{\citenamefont{Kuzmany}(2009)}]{Kuzmany2009}
\bibinfo{author}{\bibfnamefont{H.}~\bibnamefont{Kuzmany}},
  \emph{\bibinfo{title}{Ch. 6 in 'Solid-State Spectroscopy'}}
  (\bibinfo{publisher}{Springer-Verlag}, \bibinfo{year}{2009}).

\bibitem[{\citenamefont{Moon et~al.}(2010)\citenamefont{Moon, Shin, Parker,
  Choi, Mazin, Lee, Kim, Sung, Cho, Khim et~al.}}]{Moon2010a}
\bibinfo{author}{\bibfnamefont{S.~J.} \bibnamefont{Moon}},
  \bibinfo{author}{\bibfnamefont{J.~H.} \bibnamefont{Shin}},
  \bibinfo{author}{\bibfnamefont{D.}~\bibnamefont{Parker}},
  \bibinfo{author}{\bibfnamefont{W.~S.} \bibnamefont{Choi}},
  \bibinfo{author}{\bibfnamefont{I.~I.} \bibnamefont{Mazin}},
  \bibinfo{author}{\bibfnamefont{Y.~S.} \bibnamefont{Lee}},
  \bibinfo{author}{\bibfnamefont{J.~Y.} \bibnamefont{Kim}},
  \bibinfo{author}{\bibfnamefont{N.~H.} \bibnamefont{Sung}},
  \bibinfo{author}{\bibfnamefont{B.~K.} \bibnamefont{Cho}},
  \bibinfo{author}{\bibfnamefont{S.~H.} \bibnamefont{Khim}},
  \bibnamefont{et~al.}, \bibinfo{journal}{Phys. Rev. B}
  \textbf{\bibinfo{volume}{81}}, \bibinfo{pages}{205114}
  (\bibinfo{year}{2010}).

\bibitem[{\citenamefont{Wu et~al.}(2010)\citenamefont{Wu, Barisic, Kallina,
  Faridian, Gorshunov, Drichko, Li, Lin, Cao, Xu et~al.}}]{Wu2010}
\bibinfo{author}{\bibfnamefont{D.}~\bibnamefont{Wu}},
  \bibinfo{author}{\bibfnamefont{N.}~\bibnamefont{Barisic}},
  \bibinfo{author}{\bibfnamefont{P.}~\bibnamefont{Kallina}},
  \bibinfo{author}{\bibfnamefont{A.}~\bibnamefont{Faridian}},
  \bibinfo{author}{\bibfnamefont{B.}~\bibnamefont{Gorshunov}},
  \bibinfo{author}{\bibfnamefont{N.}~\bibnamefont{Drichko}},
  \bibinfo{author}{\bibfnamefont{L.}~\bibnamefont{Li}},
  \bibinfo{author}{\bibfnamefont{X.}~\bibnamefont{Lin}},
  \bibinfo{author}{\bibfnamefont{G.~H.} \bibnamefont{Cao}},
  \bibinfo{author}{\bibfnamefont{Z.~A.} \bibnamefont{Xu}},
  \bibnamefont{et~al.}, \bibinfo{journal}{Phys. Rev. B}
  \textbf{\bibinfo{volume}{81}}, \bibinfo{pages}{100512(R)}
  (\bibinfo{year}{2010}).

\bibitem[{\citenamefont{Mirri et~al.}(2012)\citenamefont{Mirri, Calvani,
  Vitucci, Peruchhi, Yeh, Wu, and Lupi}}]{Mirri2012}
\bibinfo{author}{\bibfnamefont{C.}~\bibnamefont{Mirri}},
  \bibinfo{author}{\bibfnamefont{P.}~\bibnamefont{Calvani}},
  \bibinfo{author}{\bibfnamefont{F.~M.} \bibnamefont{Vitucci}},
  \bibinfo{author}{\bibfnamefont{A.}~\bibnamefont{Peruchhi}},
  \bibinfo{author}{\bibfnamefont{K.~W.} \bibnamefont{Yeh}},
  \bibinfo{author}{\bibfnamefont{M.~K.} \bibnamefont{Wu}}, \bibnamefont{and}
  \bibinfo{author}{\bibfnamefont{S.}~\bibnamefont{Lupi}},
  \bibinfo{journal}{Supercond. Sci. Technol.} \textbf{\bibinfo{volume}{25}},
  \bibinfo{pages}{045002} (\bibinfo{year}{2012}).

\bibitem[{\citenamefont{Boris et~al.}(2009)\citenamefont{Boris, Kovaleva, Seo,
  Kim, Popovich, Matiks, Kremer, and Keimer}}]{Boris2009}
\bibinfo{author}{\bibfnamefont{A.~V.} \bibnamefont{Boris}},
  \bibinfo{author}{\bibfnamefont{N.~N.} \bibnamefont{Kovaleva}},
  \bibinfo{author}{\bibfnamefont{S.}~\bibnamefont{Seo}},
  \bibinfo{author}{\bibfnamefont{J.~S.} \bibnamefont{Kim}},
  \bibinfo{author}{\bibfnamefont{P.}~\bibnamefont{Popovich}},
  \bibinfo{author}{\bibfnamefont{Y.}~\bibnamefont{Matiks}},
  \bibinfo{author}{\bibfnamefont{R.~K.} \bibnamefont{Kremer}},
  \bibnamefont{and} \bibinfo{author}{\bibfnamefont{B.}~\bibnamefont{Keimer}},
  \bibinfo{journal}{Phys. Rev. Lett.} \textbf{\bibinfo{volume}{102}},
  \bibinfo{pages}{027001} (\bibinfo{year}{2009}).

\bibitem[{\citenamefont{Subedi et~al.}(2008)\citenamefont{Subedi, Zhang, Singh,
  and Du}}]{Subedi2008}
\bibinfo{author}{\bibfnamefont{A.}~\bibnamefont{Subedi}},
  \bibinfo{author}{\bibfnamefont{L.}~\bibnamefont{Zhang}},
  \bibinfo{author}{\bibfnamefont{D.~J.} \bibnamefont{Singh}}, \bibnamefont{and}
  \bibinfo{author}{\bibfnamefont{M.~H.} \bibnamefont{Du}},
  \bibinfo{journal}{Phys. Rev. B} \textbf{\bibinfo{volume}{78}},
  \bibinfo{pages}{134514} (\bibinfo{year}{2008}).

\bibitem[{\citenamefont{Charnukha et~al.}(2011)\citenamefont{Charnukha,
  Popovich, Matiks, Sun, Lin, Yaresko, Keimer, and Boris}}]{Charnukha2011}
\bibinfo{author}{\bibfnamefont{C.}~\bibnamefont{Charnukha}},
  \bibinfo{author}{\bibfnamefont{P.}~\bibnamefont{Popovich}},
  \bibinfo{author}{\bibfnamefont{Y.}~\bibnamefont{Matiks}},
  \bibinfo{author}{\bibfnamefont{D.~L.} \bibnamefont{Sun}},
  \bibinfo{author}{\bibfnamefont{C.~T.} \bibnamefont{Lin}},
  \bibinfo{author}{\bibfnamefont{A.~N.} \bibnamefont{Yaresko}},
  \bibinfo{author}{\bibfnamefont{B.}~\bibnamefont{Keimer}}, \bibnamefont{and}
  \bibinfo{author}{\bibfnamefont{A.~V.} \bibnamefont{Boris}},
  \bibinfo{journal}{Nature Communications} \textbf{\bibinfo{volume}{2}},
  \bibinfo{pages}{219} (\bibinfo{year}{2011}).

\end{thebibliography}
\end{document}